\documentclass[twocolumn,aps,showpacs,showkeys,superscriptaddress,floatfix]{revtex4}
\usepackage{graphicx}
\begin{document}
\title{Emission and its back-reaction accompanying electron motion in relativistically strong
and QED-strong pulsed laser fields} 
\author{Igor V. Sokolov}
\email{igorsok@umich.edu}
\affiliation{Space Physics Research Laboratory, University of Michigan, Ann
Arbor, MI 48109 }
\author{John A. Nees}
\affiliation{Center for Ultrafast Optical Science and FOCUS Center, University
of Michigan, Ann Arbor, MI 48109}
\author{Victor P. Yanovsky}
\affiliation{Center for Ultrafast Optical Science and FOCUS Center, University
of Michigan, Ann Arbor, MI 48109}
\author{Natalia M. Naumova}
\affiliation{Laboratoire d'Optique Appliqu\'{e}e, 
UMR 7639 ENSTA, Ecole Polytechnique, CNRS, 91761 Palaiseau, France}
\author{G\'{e}rard A. Mourou}
\affiliation{Institut de la Lumi\`{e}re Extr\^{e}me,  
UMS 3205 ENSTA, Ecole Polytechnique, CNRS, 91761 Palaiseau, France}
\date{\today}
\begin{abstract}
The emission from an electron in the field of a relativistically strong 
laser pulse is analyzed. At 
pulse intensities of 
$J\ge2\cdot 10^{22}\ {\rm W/cm^2}$ the 
emission from counter-propagating electrons is modified by the 
effects of Quantum ElectroDynamics (QED), as long as the electron energy is 
sufficiently high: ${\cal E}\ge 1\ {\rm GeV}$.  The radiation 
force experienced by an electron is for the first time derived from the QED principles 
and its applicability range is extended towards the QED-strong
fields.
\end{abstract}

\pacs{
52.38.-r Laser-plasma interactions,
41.60.-m Radiation by moving charges, 
52.38.Ph X-ray, gamma-ray, and particle generation }
\keywords{Thomson-Compton effect, Lorentz-Abraham-Dirac equation, 
radiation back-reaction}

\maketitle
\section{Introduction}
\subsection{QED strong fields} 
In Quantum ElectroDynamics (QED) an electric field, $E$, should be treated as  
strong if it exceeds the Schwinger limit: $E\ge E_S$, where
$$
E_S=\frac{m_ec^2}{|e|\lambdabar_C}
$$
(see \cite{schw}). Such field is potentially capable of separating a virtual 
electron-positron pair providing an energy, which exceeds the electron rest 
mass energy, $m_ec^2$, to a charge, $e=-|e|$, over an acceleration length as small 
as the Compton wavelength,
$$
\lambdabar_C=\frac\hbar{m_ec}\approx 3.9\cdot10^{-11}{\rm cm}.
$$
Spatial scales associated with 
the field 
should be greater than $\lambdabar_C$. 

Typical effects in QED strong fields are: 
electron-positron pair creation from high-energy photons, high-energy photon 
emission from electrons or positrons and electron-positron cascade 
development (see \cite{Mark}- \cite{kb}) as the result of the first two processes. 

Less typical and often forbidden by conservation laws is direct pair
separation from vacuum. This effect may only be significant if the
field invariants as defined in \cite{ll}, $F_1=({\bf B}\cdot{\bf E})$, $F_2=E^2-B^2$,
are large enough. Indeed, the considerations relating to pair creation
are applicable only in
the frame of reference in which $B=0$ or ${\bf B}\|{\bf E}$. The
electric field in this frame of reference, 
$E_0^2=F_2/2+\sqrt{F^2_1+F^2_2/4}$,
exceeds the Schwinger
limit only if the field invariants are sufficiently large.

Here  
the case of {\it weak} field
invariants is considered:
\begin{equation}\label{fielditself}
|F_1|\ll E_S^2, \qquad |F_2|\ll E_S^2,
\end{equation}     
and any corrections of the order of $F_1/E_S^2,\,F_2/E_S^2$ are
neglected (see \cite{dep} about such corrections).
So, neither the cases when the field {\it itself} is too strong, nor
the cases when its 
spatial scale is too short
are considered here. Below, 
the
term 'strong field' is only applied to the field experienced by a particle 
(electron or positron).

Particularly, a QED-strong electric field, 
\begin{equation}\label{forstar}
E_{0}=\frac{|{\bf p}\times{\bf B}|}{m_ec},
\end{equation}
may be exerted on 
relativistic charged particles with  
momentum, ${\bf p}$, gyrating in
the strong magnetic field, ${\bf B}$, of a 
neutron star, as the result of the Lorentz transformation of the 
electromagnetic field.  The field as in Eq.(\ref{forstar}) may exceed
the Schwinger limit, as long as  $|{\bf p}|\gg m_ec$ and$/$or the
magnetic field is strong enough.
 
\subsection{QED-strong laser fields}
In a laboratory experiment 
QED-strong fields may be created in 
the
focus of an ultra-bright laser.  
Consider QED-effects in  
a relativistically strong pulsed field \cite{Mark}:  
\begin{equation}\label{eq:strong}
\sqrt{{\bf a}^2}\gg1,\qquad{\bf a}=\frac{e{\bf A}}{m_ec^2},
\end{equation} 
${\bf A}$ being the vector potential of the wave. 
In the laboratory frame of reference the electric field is not QED-strong for 
achieved laser intensities, $J\sim10^{22}\ {\rm W/cm^2}$ \cite{1022}, and 
even for the $J\sim10^{25}\ {\rm W/cm^2}$ intensity projected
\cite{ELI}. Moreover, both field invariants vanish for 1D waves,  
reducing
the probability of 
direct pair creation from
vacuum  
by
virtue of the
laser field's 
proximity to 1D wave.

Nonetheless, a counter-propagating particle in a 1D  wave, ${\bf
a}(\xi),\,\xi=\omega t-({\bf k}\cdot{\bf x})$, may experience a QED-strong
field,  
$E_0=|d{\bf A}/d\xi|\omega({\cal E}-p_\|)/c$, because the laser
frequency, 
$\omega$, is Doppler
upshifted in the frame of reference co-moving with the electron. 
Herewith the electron dimensionless energy, 
${\cal E}$, and
its momentum are related to $m_ec^2$, and $m_ec$ correspondingly,
and subscript $\|$ herewith denotes the vector projection on the direction of the
wave propagation.  The Lorentz-transformed field exceeds the Schwinger limit, 
if 
\begin{equation}\label{QED-strong}
\chi\sim E_0/E_S
=\frac{\lambdabar_C}{\lambdabar}({\cal
 E}-p_\|)\left|\frac{d{\bf a}}{d\xi}\right|\gg1,
\end{equation}
where $\lambdabar=c/\omega$. Note, that the 
above-mentioned restriction on the field spatial scale  
is here assumed to be fulfilled for
the 
{\it upshifted} wave
frequency:
\begin{equation}\label{eq:boundfreq2}
\omega({\cal E}-p_\|)\ll c/\lambdabar_C.
\end{equation}
Nevertheless, the condition as in (\ref{QED-strong}) may 
be fulfilled
as
long as the field is strong enough. 
Numerical values of the parameter, $\chi$, may be
conveniently expressed in terms of the local instantaneous (not
time-average!) intensity of the laser
wave, $J$:
$$
\chi=\frac32 
\frac{\lambdabar_C}{\lambdabar}({\cal
  E}-p_\|)\left|\frac{d{\bf a}}{d\xi}\right|
\approx 0.7\frac{({\cal
  E}-p_\|)}{10^3}\sqrt{\frac{J}{10^{23}[{\rm W/cm}^2]}},
$$
the choice of the numerical factor of $3/2$ is explained below. 
For a
counter-propagating electron of 
energy 
$\sim 1$ GeV, that is for 
$({\cal E}-p_\|)\sim 4\cdot 10^3$, the QED-strength parameter is
greater than one even with the laser intensities already achieved.

The condition of $\chi>1$ also separates 
the parameter range of the Compton effect from that of the Thomson effect, 
under the condition of Eq.(\ref{eq:strong}). 
The distinctive feature of the Compton effect is  
an electron recoil, which is significant, 
if a typical emitted photon energy, $\hbar\omega_c$, is comparable with the electron energy 
\cite{kogaetal}. Their ratio, $\chi=\lambdabar_C\omega_c/(c{\cal E})$, 
equals $\chi$ as defined in Eq.
(\ref{eq:chifirst}) with the proper numerical factors 
(cf Eq.(\ref{eq:omegaccl})). It should be noted, however, that under
the conditions  
discussed 
in (\ref{eq:strong},\ref{eq:boundfreq2}) the Compton
effect drastically changes.  
\subsection{Classical radiation loss rate as an input
  parameter for QED}
Radiation processes in QED-strong fields  
are entirely controlled by the local value of 
$E_0$  (this statement may be found in \cite{lp}, \S101). 
A good signature for $E_0$ is 
the radiation loss rate
of a charge, as introduced in
the classical electrodynamics:   
\begin{equation}
I_{\rm cl}(E_0)=\frac{2e^4E^2_0}{3m_e^2c^3}=-\frac{2e^2f^\mu f_\nu}{3m_e^2c^3}, 
\end{equation}
which is a Lorentz invariant. Therefore, it may be expressed in any frame of
reference in terms of a 4-square of the Lorentz 4-force,
$f^\mu=(f^0,{\bf f})={\cal E}({\bf f}^{(3)}\cdot{\bf v}/c,{\bf
  f}^{(3)})$, where
${\bf v}$ is the velocity vector and:
$$
{\bf f}^{(3)}=e{\bf E}+\frac1c[{\bf v}\times{\bf B}],
$$
$$ f^0=e{\bf
E}\cdot{\bf p},\qquad {\bf f} = e{\cal E}{\bf E}+e[{\bf p}\times{\bf B}]).
$$ 
So,  
the QED-strength of the 
field may be determined in  
evaluating $I_{\rm cl}$ and  
its ratio to 
$I_C=I_{\rm cl}(2E_S/3)$: 
\begin{equation}\label{eq:chifirst}
\chi= \sqrt{\frac{I_{\rm cl}}{I_C}},\qquad
I_C=\frac{8e^2c}{27\lambdabar_C^2}.
\end{equation} 
If $\chi\ge 1$ then  
the actual radiation loss rate differs 
from $I_{\rm cl}$, however, it may be re-calculated using $I_{\rm cl}$
as a sole input parameter.
     
\subsection{Possible realization of QED-strong fields in laboratory experiments} 
The first experiments which demonstrated QED effects in a laser field
were fulfilled using an electron beam of 
energy 
$\approx 46.6$ GeV (see
\cite{bb}), which interacted with a counter-propagating terawatt laser
pulse of 
intensity 
$J\sim10^{18}{\rm W/cm}^2$. A reasonably high
value of $\chi\approx 0.4$ had been achieved,  however, the
laser field was not relativistically strong with $|{\bf a}|\le 1$. The
high value of $\chi$ had been achieved at the cost of very high energy
of the upshifted laser wave: the transformed photon energy amounted to
$\sim 0.1$ MeV, 
which is not small as compared to the electron rest mass
energy, $m_ec^2\approx 0.51$ MeV. It could be interesting to upgrade
this experiment towards the highest achievable laser intensities 
$\ge 2\cdot 10^{22}$ with the use of a wakefield-accelerated beam of
electrons of  
energy 
$\sim1$ GeV (see \cite{heg}). First, the
2-3 times larger value of the QED parameter, $\chi$, may be achieved
with the exponentially increased probability for pair creation. Second,
such experiment would be highly relevant to the processes which will
occur in the course of laser-plasma interaction at even stronger laser
intensities.
   
Indeed, counter-propagating electrons can be generated while a laser 
pulse is interacting with a solid target. For this reason, the radiation effects in the course 
of laser-plasma interaction are widely investigated ((see
\cite{kogaetal,lau03}). With future progress in laser technology and
by achieving intensities of $J\sim 10^{23} {\rm W/cm}^2$  
laser-plasma interactions  
will be strongly
modified by QED effects, 
so that the capability to model these effects now is of interest. 
\subsection{Radiation back-reaction}
The principle matter in this paper is an account of the radiation
back-reaction acting on a charged particle. 
The radiation losses reduce the particle energy, affecting 
both the particle motion and the radiation losses themselves.

This effect can be consistently described 
by solving the dynamical equation which appears to reduce to the modified
Lorentz-Abraham-Dirac equation  
as derived in \cite{mine},\cite{our}. The new element 
here
is that in this equation
the radiation back-reaction on the electron motion should be expressed in 
terms of the emission probability, while applied to QED-strong fields. 

In section II the emission from an electron in the field of a
relativistically strong wave 
is discussed within the framework of
classical electrodynamics. The transition from
the vector amplitude of emission as described in \cite{jack} to the 
instantaneous spectrum of emission is treated in terms of the 
{\it formation time}, 
a concept, which is not often used in classical 
electrodynamics.  For QED-strong laser pulses the calculation of the
emission probability 
is given in Section III. The radiation processes
in QED-strong fields 
appear 
to be reducible to 
a frequency downshift
in the classical vector amplitude of emission resulting from the
electron recoil, accompanied 
by a contribution to emission  
associated with the magnetic moment of electron.   
The radiation effect on the electron motion in strong fields is
discussed in Section IV. 
The 
Conclusion
summarizes 
the results  
and 
discusses 
future prospectives.
\section{Emission in relativistically strong fields}
In this section 
QED effects are not yet considered, but the
electromagnetic wave field is assumed to be relativistically strong.
Angular and frequency 
distributions 
of 
electron emission  
are discussed. 
The goal is to
establish a connection between the methods usually applied to calculate
emission in weaker fields,  
on
one hand, and the conceptually different QED approach
on the other. 
For relativistically
strong laser fields, even though QED effects do not yet come into
a power, still some concepts of the QED emission theory
appear to be applicable  and
useful, among them are the formation time of emission and
instantaneous spectrum of emission.  

In weaker fields, especially for 
the particular case of 
a harmonic wave, the emitted power is given by an integral over many
periods of the wave. This standard approach, however, may become
meaningless as applied to the ultra-strong laser
pulses, for many reasons. These pulses may be so short that they cannot 
be thought of as 
harmonic waves. 
Their fields may be strong enough to force an
electron to 
expend its energy on radiation faster than a single
wave period. However, an even more important point is that the radiation
loss rate and even the spectrum of radiation is no longer an integral 
characteristic of the particle motion through a number of wave periods:
a local dependence of emission on both particle and field
characteristics is typical for the strong fields.
\subsection{Transformed space-time}
A 
method facilitating 
many derivations 
involves the introduction of a specific time-space coordinate frame.   
Consider  
a 1D wave field  
taken in the Lorentz calibration: 
$$
a^\mu=a^\mu(\xi),\qquad \xi=(k\cdot x),\qquad(k\cdot a)=0,
$$ 
$a^\mu=(0,{\bf a})$, $k^\mu$ and $x^\mu$ being the
4-vectors of the 
potential, the wave and the coordinates. Herewith
the 4-dot-product is introduced in a 
usual manner: 
$$(k\cdot x)=k^\mu x_\mu = \omega t-({\bf k}\cdot{\bf x})$$ 
etc. Space-like 3-vectors (i.e., the first to the third components of
a 4-vector)  in contrast with 4-vectors are denoted in bold,
4-indices are denoted with Greek letters. Note that a 
metric signature $(+,-,-,-)$ is used, therefore, for
space-like vectors the 3D scalar product and 4-dot-product have
opposite signs, particularly:
$$
\left(\frac{d{\bf a}}{d\xi}\right)^2=
-\left(\frac{da}{d\xi}\right)^2\ge 0.
$$    

Introduce a 
Transformed Space-Time (TST) :  
$$
x^{0,1}=(ct\mp x_\|)/\sqrt{2},\qquad x^{2,3}={\bf x}_\perp,
$$ 
subscript $\perp$ denoting the vector components 
orthogonal to ${\bf k}$. The properties of the TST provide a convenient
description for the classical motion of an electron in the 1D wave
field. First, note, that
$$
dx^0=\frac{\lambdabar d\xi}{\sqrt{2}}, \qquad
p^0=\frac{\lambdabar(k\cdot p)}{\sqrt{2}}, \qquad (p\cdot k)= 
\frac{{\cal E}-p_\|}{\lambdabar}.
$$ 
Second,  
the generalized momentum components,  
$p^0$ and ${\bf p}_{\perp 0}={\bf p}_\perp+{\bf a}$,  
are conserved. Third, the metric tensor in the TST is:
$$
G^{01}=G^{10}=1,\qquad G^{22}=G^{33}=-1, \qquad G^{\mu\nu}=G_{\mu\nu}. 
$$
Finally, the identity, ${\cal E}^2=p^2+1$, being expanded in the TST metric, gives: 
$$
p^1=\frac{1+{\bf p}_\perp^2}{2p^0}=
\frac{1+({\bf p}_{\perp 0}-{\bf a})^2}{\sqrt{2}\lambdabar(k\cdot p)}.
$$  
The classical
radiation loss rate is found by virtue of expanding the Lorentz force
squared in the TST:
\begin{equation}\label{eq:classicintense}
I_{\rm cl}=-\frac{2e^2}{3c}\frac{(f\cdot f)}{m_e^2c^2}=\frac{2e^2c(k\cdot
 p)^2}{3}\left(\frac{d{\bf a}}{d\xi}\right)^2.
\end{equation}
The derivative over $x^0$ or, the same, over $\xi$ is conveniently
related to the derivative over the proper time for electron: 
\begin{equation}
\frac{d}{d\tau}={\cal E}\left[\frac\partial{\partial t}+({\bf
    v}\cdot\frac\partial{\partial {\bf x}})\right]=c(k\cdot
p)\frac{d}{d\xi}.
\end{equation}
\subsection{Classical trajectory and momenta retarded product}
Many characteristics of emission may be expressed in terms of 
the 
relationship between 
the
4-momenta of 
the
electron
at different 
instants:
\begin{equation}\label{eq:classic}
p^\mu(\xi)=p^\mu(\xi^\prime)-\delta a^\mu+
\frac{2(p(\xi^\prime)\cdot \delta a)-(\delta a)^2}{2(k\cdot p)}k^\mu,
\end{equation}
where 
$$\delta a^\mu=a^\mu(\xi)-a^\mu(\xi^\prime).$$
As a 
consequence 
from Eq.(\ref{eq:classic}), one can obtain the expression for the {\it Momenta Retarded Product} (MRP):
\begin{equation}\label{eq:pp}
(p(\xi)\cdot p(\xi^\prime))=1-\frac{(\delta a)^2}2=1+\frac{(\delta{\bf a})^2}2,
\end{equation}

Note, that the MRP is given by Eq.(\ref{eq:pp}) for an  
arbitrary difference 
between $\xi$ and $\xi^\prime$, but only for 
the particular case of the 1D
wave field. However the limit of this formula 
as 
$|\xi-\xi^\prime|\rightarrow 0$, which is as follows:
$$
(p(\xi)\cdot p(\xi^\prime))|_{|\xi-\xi^\prime|\rightarrow 0}\approx 
1+ \frac12(\xi-\xi^\prime)^2\left|\frac{d{\bf a}}{d\xi}\right|^2
$$
or, in terms of the 
MRP in the proper time, $\tau$:
\begin{equation}\label{eq:gencovariance}
(p(\tau)\cdot p(\tau+\delta\tau))=1-(\delta\tau)^2\frac{(f\cdot f)}{2m_e^2c^2},
\end{equation}
has a much wider range of applicability. Eq.(\ref{eq:gencovariance}) is
derived from the equation of
motion:
$$
\frac{dp^\mu}{d\tau}=\frac{f^\mu}{m_ec},
$$
using the identities: 
$$
(p(\tau)\cdot p(\tau))=1,\qquad (p(\tau)\cdot f(\tau))=0,
$$
$$
\frac{d(p(\tau)\cdot p(\tau+\delta\tau))}{d(\delta\tau)}=-\delta\tau\left(\frac{dp}{d\tau}\cdot\frac{dp}{d\tau}\right)+O((\delta\tau)^2).
$$
\subsection{Vector amplitude of emission in classical electrodynamics}
In Ref.\cite{jack}
the frequency spectrum and angular distribution, 
$dR_{\rm cl}/(d\omega^\prime d{\bf n})$,
of the radiation energy, $dR_{\rm cl}$, emitted by an electron and related to the interval of
frequency, $d\omega^\prime$, and to the element of solid angle, $d{\bf
  n}$, for a 
polarization vector, ${\bf l}$, is described with the following formula:
\begin{equation}\label{eq:Jackson}
\frac{dR_{\rm cl}}
{d\omega^\prime d{\bf n}}=\frac{(\omega^\prime)^2}{4\pi^2c}\left|({\bf
  A}_{\rm cl}(\omega^\prime)\cdot {\bf l}^*)\right|^2.
\end{equation}
Here the superscript asterisk means the complex conjugation and the vector amplitude of emission, ${\bf
  A}_{\rm cl}(\omega^\prime)$, is given by the following equation:
$$
{\bf A}_{\rm cl}(\omega^\prime,{\bf n})=\frac{e}c\int_{-\infty}^{+\infty}{{\bf
    v}(t)\exp\left\{i\omega^\prime[t-\frac{({\bf n}\cdot{\bf r}(t))}c]\right\}dt},
$$
see Eq.(14.67) in Ref.\cite{jack} followed by the discussion of the way
to account for a polarization. 
The use of the same notation, ${\bf A}$, both for the emission
vector amplitude and for the vector potential should not mislead the
reader. Recall, that the emission vector amplitude is closely related to
the Fourier-transformed vector potential in the far-field zone of
emission. 

Introduce a 4-vector amplitude of emission, 
$$
A^\mu_{\rm cl}(\omega^\prime,{\bf n})=e
\int_{-\infty}^{+\infty}{
p^\mu(\tau)
\exp\left[
ic\int^\tau{(k^\prime\cdot p(\tau^\prime))d\tau^\prime}\right]d\tau},
$$ 
which is expressed in terms of the proper time for the electron and its
(dimensionless) 4-momentum. As long as $(k^\prime\cdot
p)/c$ is a frequency of the emitted photon in the frame of
reference  
co-moving with the electron, the 4-vector amplitude
is the Fourier integral of the electron 4-momentum with
the Lorentz-modified frequency. 

Note the following properties of the
4-vector amplitude. First, its space-like vector components, which
are perpendicular to the wave vector of the emitted photon, coincide
with those for 3-vector amplitude,  hence, they quantify the
polarization properties of the emission for two different
polarizations. Second, the dot-product, $(A_{\rm cl}\cdot k^\prime)$,
vanishes as being the integral of a perfect time derivative. Now
construct the dot-product, $(A_{\rm cl}\cdot A^*_{\rm cl})$ and expand it
in the TST, which is formulated in terms of the {\it emitted} wave: 
$$
(A_{\rm cl}\cdot A^*_{\rm cl})=A^0_{\rm cl}(A^1_{\rm cl})^* +A^1_{\rm
    cl}(A^0_{\rm cl})^* 
-|A^2_{\rm cl}|^2 - |A^3_{\rm cl}|^2. 
$$
From the above properties of the 4-vector amplitude, the first two
terms vanish identically as $A^0_{\rm cl}\propto ( A_{\rm
  cl}\cdot k^\prime)=0$. The other two terms as taken with the proper factor give the
emitted energy summed up over polarizations, therefore, the latter sum
may be expressed as follows:
\begin{equation}\label{eq:sumdr}
\sum_{\bf l}{\frac{dR_{\rm cl}}{d\omega^\prime d{\bf n}}}=-
\frac{(\omega^\prime)^2}{4\pi^2c}(A_{\rm cl}\cdot A^*_{\rm cl}).
\end{equation}
Now introduce the radiation loss rate, $dI_{\rm cl}/(d\omega^\prime
d{\bf n})$ related to the unit of time, the element of a solid
angle and the frequency interval. Its connection to $dR_{\rm
  cl}/(d\omega^\prime d{\bf n})$ is evident:
$$
\sum_{\bf l}
{\frac{dR_{\rm cl} }
{d\omega^\prime d{\bf n} } 
}=
\int_{-\infty}^{+\infty}{
\frac{dI_{\rm cl}(t)}
{d\omega^\prime d{\bf n} } dt}=
\int_{-\infty}^{+\infty}{\frac{dI_{\rm cl}(\tau)}{d\omega^\prime
d{\bf n}}{\cal E}(\tau)d\tau}.
$$
In Eq.(\ref{eq:sumdr}) the dot-product of the
4-vector amplitudes,
$(A_{\rm cl}\cdot A^*_{\rm cl})$ is in fact the product of two
integrals over $d\tau$, which
can be represented as the double integral, over,
say, $d\tau_1d\tau_2$. Transform the
integration variables in this double integral, by introducing $\tau=(\tau_1+\tau_2)/2$, 
$\theta=\tau_1-\tau_2$. 
The 
spectral and angular distribution of the radiation loss rate
may be expressed in terms of the Fourier integral of the  
MRP:
$$
\frac{dI_{\rm cl}(\tau)}{d\omega^\prime
d{\bf n}}=-
\frac{e^2(\omega^\prime)^2}
{4\pi^2c{\cal E}(\tau)}
\int_{-\infty}^{+\infty}
(p(\tau+\frac\theta2)\cdot p(\tau-\frac\theta2) )
\times
$$
$$
\times
\exp\left[
ic
\int_{\tau-\theta/2}^{\tau+\theta/2}
{(k^\prime\cdot p(\tau^\prime))d\tau^\prime}\right]d\theta.
$$
\subsection{Frequency spectrum and formation time}
The specific feature of the particle relativistic motion in strong
laser fields is that the main contribution to the above integral comes
from a brief time interval with small values of $\theta$. A closely related
point is that the emitted radiation is abruptly beamed about the 
direction of the velocity vector, ${\bf p}(\tau)/|{\bf p}(\tau)|$. 
Therefore, in the following expansion of the frequency in the co-moving
frame,
$$
(k^\prime\cdot p(\tau^\prime))=
\left([k^\prime-\frac{\omega^\prime}{c{\cal E}(\tau)}p(\tau)]\cdot
p(\tau^\prime)\right)+\frac{\omega^\prime}{c{\cal E}}\left(p(\tau)
\cdot p(\tau^\prime)\right),
$$  
in the first dot-product one may approximate
$p^\mu(\tau^\prime)\approx p^\mu(\tau)$. Indeed, for an ultrarelativistic
electron and a photon, which both propagate in the same direction, the
difference between 
$p^\mu/{\cal E}$ and
$c(k^\prime)^\mu/\omega^\prime$ is already small, therefore in the
second multiplier of the dot-product the small difference,
$p^\mu(\tau)-p^\mu(\tau^\prime)$, may be neglected: 
$$
c\int_{\tau-\theta/2}^{\tau+\theta/2}{(k^\prime\cdot
  p(\tau^\prime))d\tau^\prime}\approx
\theta c (k^\prime\cdot p(\tau))+
$$
$$
+\frac{\omega^\prime}{\cal E}\int_{\tau-\theta/2}^{\tau+\theta/2}{\left[\left(p(\tau)\cdot
      p(\tau^\prime)\right)-1\right]d\tau^\prime}.
$$  
Now the only angle-dependent multiplier is $\exp[-i\theta c
  (k^\prime\cdot p(\tau))]$. For simplicity, the angular spectrum of
emission can be approximated with the Dirac function:
$$
\frac{dI_{\rm cl}(\tau)}{d\omega^\prime
d{\bf n}}=\delta^2\left({\bf
  n}-\frac{\bf p}{|{\bf p}|}\right)\frac{dI_{\rm
    cl}(\tau)}{d\omega^\prime},
$$
and with the use of the formula (see \S90 in Ref.(\cite{lp}),
$$
\int{\exp[i\theta c
  (k^\prime\cdot p(\tau))]d{\bf
    n}}=
\frac{2\pi i}
{\omega^\prime{\cal E}(\tau)\theta}\exp\left(\frac{i\omega\theta}{2{\cal E}(\tau)}\right),
$$
the following expression may be found for the frequency spectrum of
emission:
$$
\frac{dI_{\rm cl}(\tau)}{d\omega^\prime}=
\frac{e^2\omega^\prime}
{2\pi c{\cal E}^2(\tau)}
\int_{-\infty}^{+\infty}
\frac1\theta(p(\tau+\frac\theta2)\cdot p(\tau-\frac\theta2) )
\times
$$
$$
\times
\sin\left\{
\frac{\omega^\prime}{{\cal E}(\tau)}
\left[\frac\theta2+
\int_{\tau-\theta/2}^{\tau+\theta/2}
{[(p(\tau)\cdot p(\tau^\prime))-1]d\tau^\prime}
\right]
\right\}
d\theta.
$$
Thus, the frequency spectrum of emission is entirely determined by the
MRP, which is a scalar Lorentz-invariant funcion
of the proper time. Both the fore-exponential factor and the
argument of the exponential function depend on the mentioned
MRP. Therefore, both the spectral composition of the MRP
and its magnitude may be of importance. Their relative role is controlled by
the ratio of the frequency of the electron motion, $\omega_0$, to the
acceleration magnitude, both being  determined in the
co-moving frame of reference. Here the field is assumed to be so strong, that the
acceleration it causes plays the dominant role, i.e. the following
inequality is claimed:
\begin{equation}\label{eq:highac}
-\frac{(f\cdot f)}{m_e^2 c^2}\gg \omega_0^2.
\end{equation}
Under these circumstances, the integral determining the emission
spectrum
is calculated by virtue of the displacement of the integration contour
in the plane of the complex variable, $\theta$, so that the deformed
contour passes through the point
of a {\it stationary phase}, $\theta_{\rm st}$. In this 'saddle' point the phase
gradient turns to zero:
$$
\frac{d}{d\theta}\left[\frac\theta2+\int_{\tau-\theta/2}^{\tau+\theta/2}
{[(p(\tau)\cdot p(\tau^\prime))-1]d\tau^\prime}\right]=0.
$$
The larger the acceleration becomes, the closer the
stationary phase point, 
$\theta_{\rm st}$ draws
to the real axis, and, hence,
the shorter the time interval becomes, 
$\theta\sim \theta_{\rm f}=|\theta_{\rm st}|$, which gives the non-vanishing
contribution to the emission spectrum. The characteristic duration of
this time interval, $\theta_{\rm f}=|\theta_{\rm st}|$ is referred to
as a formation time (or coherence time - see \cite{nr},\cite{kat}). At the limit of large accelerations the
formation
time is given by the following formula:
\begin{equation}
\theta_{\rm st}
=
\pm i
\frac{2m_ec}{\sqrt{-(f\cdot f)}},\qquad \theta_{\rm f}=\frac{2m_ec}{\sqrt{-(f\cdot f)}},
\end{equation}
where the approximation for the MRP as in
Eq.(\ref{eq:gencovariance}) is applied at
$|\theta|\le \theta_{\rm f}$. With the use of
Eq.(\ref{eq:gencovariance}) the {\it universal} emission spectrum 
is obtained: 
$$
\frac{dI_{\rm cl}(\tau)}{d\omega^\prime}=
\frac{e^2\omega^\prime}
{2\pi c{\cal E}^2(\tau)}
\int_{-\infty}^{+\infty}
[\frac1\theta-\frac{(f(\tau)\cdot f(\tau))\theta}{2m_e^2c^2}]
\times
$$
$$
\times
\sin\left\{
\frac{\omega^\prime}{{\cal E}(\tau)}
\left[\frac\theta2-\frac{(f(\tau)\cdot f(\tau))\theta^3}{24m_e^2c^2}
\right]
\right\}
d\theta.
$$
The integral can be expressed in terms of the MacDonald function (= the
modified Bessel function):
$$
\frac{dI_{\rm cl}(\tau)}{d\omega^\prime}=\frac{I_{\rm
    cl}(\tau)}{\omega_c}Q_{\rm cl}(r_0),\qquad \frac{dI_{\rm cl}(\tau)}{dr_0}=I_{\rm
    cl}(\tau)Q_{\rm cl}(r_0),
$$ 
where $Q_{\rm cl}(r_0)$ is the 
unity-normalized 
spectrum of the
gyrosynchrotron emission ($\int{Q_{\rm cl}(r)dr}=1$):
\begin{equation}
Q_{\rm cl}(r_0)=
\frac{9\sqrt{3}}{8\pi}r_0
\int_{r_0}^\infty{K_{5/3}(r^\prime)dr^\prime},\qquad
r_0=\frac{\omega^\prime}{\omega_c},\label{eq:classicspectrum}
\end{equation}
and
$$
\omega_c=\frac32
   \frac{
{\cal E}(\tau)\sqrt{-(f(\tau)\cdot f(\tau))}
}
{m_ec}=
\frac32
{\cal E}(\tau)
\sqrt{
\frac{3I_{\rm cl}(\tau)c}{2e^2}},
$$
or, which is the same,
\begin{equation}\label{eq:omegaccl}
\frac{\hbar\omega_c}{m_ec^2}={\cal E}\sqrt{\frac{I_{\rm cl}(\tau)}{I_C}}={\cal E}\chi.
\end{equation}
Note, that despite all approximations, the integral over the
frequency spectrum is consistently equal to $I_{\rm cl}$.
\subsection{Implications for strong laser fields} 
As discussed, the condition ${\cal E}\gg1$ and
the inequality (\ref{eq:highac}) are both fulfilled
for an 
ultra-relativistic electron gyrating in a uniform
steady-state magnetic
field. By expressing the 4-force squared, $-(f\cdot f)=e^2{\bf
  p}_\perp^2{\bf B}^2$ (see Eq.(\ref{forstar})), and taking the gyrofrequency in the
co-moving frame, $\omega_0^2={\cal E}^2e^2{\bf B}^2/(m_e^2c^2{\cal
  E}^2)=e^2{\bf B}^2/(m_e^2c^2)$, one finds that (\ref{eq:highac}) is
fulfilled as long as ${\bf p}_\perp^2\gg1$. 

Furthermore, application to the 1D wave field
is no less straight forward.
The laser wave frequency in the comoving frame, $\omega_0=c(k\cdot p)$,
is present  
on the left hand side of Eq.(\ref{eq:boundfreq2}). The
Lorentz 4-force squared is given in Eq.(\ref{eq:classicintense}),
resulting in the following estimate for the formation time:
\begin{equation}\label{eq:thetafforwave}
\theta_{\rm f}=\frac{2}{c(k\cdot p)|d{\bf a}/d\xi|}
\end{equation} 
Now it is easy to see, that the condition, $\theta_{\rm f}\omega_0\ll1$ as in
Eq.(\ref{eq:highac}) is fulfilled in relativistically strong wave
field, at
\begin{equation}
\left|\frac{d{\bf a}}{d\xi}\right|\gg1.
\end{equation}
The formation time tends to zero as the wave amplitude tends to
infinity. The change in the electron energy within the formation time
is always  
much 
less than the particle energy, ${\cal E}m_ec^2$. Within the classical
field theory this statement follows from 
Eqs.(\ref{eq:classicintense},\ref{eq:thetafforwave}).  With an account of an extra factor
of ${\cal E}$, which arises while  
transforming 
the formation time to the laboratory
frame of reference, the relative change in energy equals:
\begin{equation}\label{eq:Stepansratio}
\frac{\theta_{\rm f}I_{\rm cl}}{m_ec^2}=2\alpha\chi,
\end{equation} 
where
$$\alpha=\frac{e^2}{\hbar c}\approx\frac1{137}$$
is  
fine structure constant. 
The ratio (\ref{eq:Stepansratio}) is much less than 
unity as long as
$\chi\le 1$. Note that in the
opposite limiting case of QED-strong field, the extra factor of $I_{\rm
  QED}/I_{\rm cl}\propto \chi^{-4/3}$ (see Fig.\ref{fig_3} below) makes 
  ratio
(\ref{eq:Stepansratio}) 
small at $\chi\gg1$ 
as well.

The same estimate of the formation time is applicable to any
relativistically strong electromagnetic field,
not 
only to 1D wave. Particularly, in the
wakefield acceleration scheme, where  
the electric
field in the wakefield of the pulse may be even larger than the
relativistically strong field in the pulse
itself. With this account, the acceleration of almost monoenergetic
electron beam by the laser pulse must be accompanied by the
gyrosynchrotron-like spectrum of emission (which are actually observed -
see \cite{Rousse}, \cite{kneip}). 
These observations 
demonstrate the general character of the
gyrosynchrotron emission spectrum (this point of view, presented in
\cite{Rousse}, may be also found in \S77 in \cite{ll}). 

\subsection{Emission within short time intervals and implications for
  numerical simulation} 
In strong fields satisfying the condition as in (\ref{eq:highac}) both
emission vector amplitude and the emission spectrum may be determined
with respect to {\it brief} time interval, $\Delta t$, which may be
much shorter than the field period. The only requirement is that this
interval should be large as compared to the formation time:
\begin{equation}\label{eq:Deltat}
\int_t^{t+\Delta t}{\frac{dt^\prime}{{\cal E}(t^\prime)}}\gg \theta_{\rm f},
\end{equation}
however, the change in the field and particle characteristics within
this time interval may be small, as long as $\Delta t\omega\ll1$. In
this case the span in the integral determining  the vector amplitude
may be chosen to be $(t,t+\Delta t)$. However, in the integral over
$d\theta$, which determines the emission spectrum, the integration
limits are much larger than the formation time, therefore, they may be
again set to $(-\infty,+\infty)$.

These 
considerations 
justify the numerical scheme for
collecting the high frequency emission as described in \cite{our}
(which does not seem different from that briefly described in
\cite{Rousse}). In addition to
calculating the electromagnetic fields on the grid using Particle-In-Cell (PIC) 
scheme, 
in 
which fields are created by a moving particle within the time
step, $\Delta t$, one can also account for the higher-frequency (subgrid)
emission spectrum, 
by calculating the instantly radiated energy, $I_{\rm
  cl}\Delta t$, and its distribution over frequency, parameterized via $I_{\rm
  cl}$. Another often used approach based on the calculation of the
vector amplitude of emission (see, e.g.,\cite{spie}) seems to be less efficient, although,
theoretically, should provide the same result. The vector
amplitude formalism, on the other hand, may be better applicable to the
cases, 
where the high-frequency emission from multiple electrons is
coherent (see \cite{habs}). 
Stemming from these considerations it is now easy to proceed to the QED approach.  
\section{Electron in QED-strong field: the emission probability} 
The emission probability in the QED-strong 1D wave field may be found in 
\S\S40,90,101 in \cite{lp}, as well as in \cite{nr},\cite{gs}. In
application to the wakefield acceleration of electrons of energy 
$\approx 1$ TeV the QED effects had been also discussed in \cite{Khok}. However, to 
simulate highly dynamical effects in pulsed fields, one needs 
a reformulated emission probability, related to short time intervals 
(not $(-\infty,+\infty)$).

Indeed, it is demonstrated above that in 
strong
fields the emission processes are essentially local functions of the
instantaneous parameters. Therefore, in QED-strong 
fields the emission
probability should be formulated in terms of the
local values of the electromagnetic field intensities, or, the way we
adopt, it may be parameterized via the classical radiation loss rate or the Lorentz 
4-force squared: $-f^\mu f_\mu\propto I_{\rm cl}$. This emission
probability is rederived here with careful attention to
consistent problem formulation and neglecting technical details.
\subsection{A QED solution of the Dirac equation} 
The Dirac equation which determines the evolution of the wave function, 
$\psi$, for a {\it non-emitting} electron in the external field, reads:
\begin{equation}\label{eq:Dirac}
\left[i \lambdabar_C\left(\gamma\cdot\frac\partial{\partial
  x}\right)-(\gamma\cdot a)\right]\psi=\psi,
\end{equation}
$\gamma^\mu$ being the Dirac $4*4$ matrices,
$(\gamma^0,\gamma^1,\gamma^2,\gamma^3)$. The relativistic dot-product
of the Dirac matrices by 4-vector, such as $(\gamma\cdot a)$, is the
linear combination of the Dirac matrices: 
$(\gamma\cdot a)=\gamma^0a^0-\gamma^1a^1-\gamma^2a^2-\gamma^3a^3$. Such
linear combination, which is also
a
$4*4$ matrix, may be multiplied
by another matrix of this kind or by 4-component bi-spinor, such as $\psi$,
following 
matrix multiplication rules. For example, $(\gamma\cdot
a)\psi$ is a bi-spinor, as
is
the matrix, $(\gamma\cdot a)$ multiplied from the right
hand side by the bi-spinor, $\psi$. 

By expanding Eq.(\ref{eq:Dirac})
in the TST it is easy to find its solution in a form of a plane
electron wave 
(the normalization coefficient $N={\rm const}$):
\begin{equation}\label{eq:volkov}
\psi=\frac{ u(p(\xi))P(\xi)}{\sqrt{N}}
\exp\left[
\frac{i
\left[
({\bf p}_{\perp0}\cdot {\bf x}_\perp)-
\frac{\lambdabar(k\cdot p)x^1}{\sqrt{2}}
\right]
}{\lambdabar_C}
\right].
\end{equation}
Here
$u(p(\xi))$ is plane wave bi-spinor amplitude, which satisfies the
system of four linear algebraic equations: 
\begin{equation}\label{bispinor}
(\gamma\cdot p(\xi))u(p(\xi))=u(p(\xi)),
\end{equation}
as well as the normalization condition:
 $\hat{u}u=2.$ 
The $\xi$-dependent momentum, $p(\xi)$, in the bi-spinor 
amplitude should be taken in accordance with Eq.(\ref{eq:classic}) as
for the classical trajectory of the electron.  
The  $\xi$-dependent phase multiplier, $P(\xi)$ is as follows:
$$
P(\xi)=\exp\left(-\frac{i}{\lambdabar_C}
\int^{\xi}{\frac{1+{\bf p}_\perp^2(\xi_2)}{2(k\cdot p)}
d\xi_2}\right),
$$
or,
\begin{equation}\label{xidependent}
P(\xi)=P(\xi^\prime)\exp\left(-\frac{i}{\lambdabar_C}
\int^{\xi}_{\xi^\prime}{\frac{1+{\bf p}_\perp^2(\xi_2)}{2(k\cdot p)}
d\xi_2}\right).
\end{equation} 
Using Eq.(\ref{eq:classic}), one can find:
\begin{equation}\label{eq:QEDpropagator}
u(p(\xi))=\left[1+\frac{(\gamma\cdot
 k)\left(\gamma\cdot [a(\xi)-a(\xi^\prime)]
\right)}{2(k\cdot p)}\right]u(p(\xi^\prime))
\end{equation}  
and verify that Eq.(\ref{eq:volkov}) satisfies the Dirac equation. The 
advantage of the 
approach used here  
as compared to the known
Volkov solution presented in \S40 in \cite{lp} is that the wave
function in Eqs.(\ref{eq:volkov}-\ref{eq:QEDpropagator})
is described in a self-contained manner within some finite time
interval, $(\xi^\prime,\xi)$ (in fact, 
this interval is assumed 
to be very short below) in terms of the local parameters of the classical 
trajectory of electrons. This 
approach is better applicable to
strong fields, in which the time interval between subsequent emission
occurrences, 
which destroys the unperturbed wave function, becomes very short.
\subsection{The matrix element for emission} 
The emission problem is formulated in the following way. The electron
motion in the strong field may be thought of as the sequence of short
intervals. Within each of these intervals the electron follows a piece
of a classical trajectory, as in Eq.(\ref{eq:classic}), and its wave
function (an electron state) is given by Eq.(\ref{eq:volkov}). The 
transition from one piece of the classical trajectory to another, or, 
the same, from one electron state
to another occurs in a probabilistic manner. The probability of this
transition, which is accompanied by a photon emission is calculated
below using the QED perturbation theory. 

The only difficulty specific
to strong pulsed fields is that the short piece of the electron
trajectory is strictly bounded in space and in time, while the QED
invariant perturbation theory is based on the 'matrix element', which
is the integral over infinite 4-volume.       
\begin{figure}
\includegraphics[scale=0.4, angle=90]{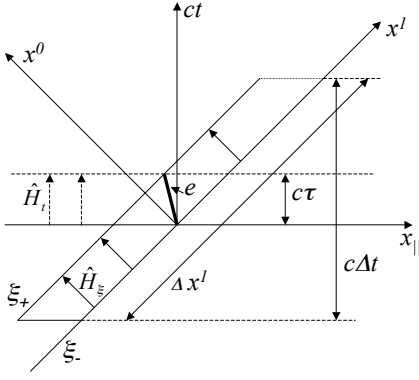} 
\caption{The volume over which to integrate the matrix element while finding 
the emission probability: in the standard scheme for the dipole emission 
(in dashed lines) and in the TST (in solid lines). Arrows show the 
direction, 
along which the Heisenberg operator advances the wave functions.
}
\label{fig_1}
\end{figure} 

To avoid this difficulty the following method is suggested, which is
analogous to the dipole emission theory as applied in TST. Introduce
domain, $\Delta^4x=(\Delta x^1*S_\perp)*\Delta x^0$, 
bounded by two hypersurfaces, $\xi=\xi_{-}$ and $\xi=\xi_+$
(see Fig.1). The difference $\xi_+-\xi_-$ 
is bounded as described below,
so that $\Delta^4x$ covers only a minor part of the pulse.   
A volume, 
$$
V=S_\perp\lambdabar(\xi_+-\xi_-)
=S_\perp\lambdabar\int_{\xi_-}^{\xi_+}{d\xi_2},$$ 
is a section of $\Delta^4x$
subtended by a line $t={\rm const}$. 

With the following choice for the normalization coefficient in Eq.(\ref{eq:volkov}): 
$$
N=2S_\perp \lambdabar\int_{\xi_-}^{\xi_+}{{\cal E}
(\xi_2)d\xi_2},
$$  
the integral of the electron density in the volume V, 
$$
\int{\hat{\psi}\gamma^0\psi dV}=
S_\perp \lambdabar\int_{\xi_-}^{\xi_+}{\hat{\psi}\gamma^0\psi
  d\xi_2},$$ 
is set to unity, i.e. there is a single electron in the volume
$V$. This statement follows from  Eq.(\ref{eq:volkov}) and the known
property of normalized bi-spinor amplitudes:
$\hat{u}\cdot\gamma^0 \cdot u=2{\cal E}$. Here the hat means the Dirac conjugation.
 
For a photon of 
wave vector, $(k^\prime)^\mu$, and polarization vector, $l^\mu$,
introduce the 
wave function:
$$
(A^\prime)^\mu=
\frac{
\exp[-i(k^\prime\cdot x)/\lambdabar_C]
     }
{
 \sqrt{N_p}
}l^\mu,
$$
or, by expanding this in the TST:
$$
(A^\prime)^\mu=
\frac{P_p(\xi)}{\sqrt{N_p}}
\exp\left[
\frac{i({\bf k}^\prime_\perp\cdot {\bf x}_\perp)}{\lambdabar_C}-
\frac{
i\lambdabar (k\cdot k^\prime) x^1}
{\sqrt{2}\lambdabar_C}
     \right]
l^\mu,$$
where:
$$P_p(\xi)=
\exp\left[-i \xi
\frac{ ({\bf k}^\prime_\perp)^2}
{2(k\cdot k^\prime)\lambdabar_C}
\right].
$$
Here the photon momentum and photon energy are related to $m_ec$ and
$m_ec^2$ correspondingly, or, equivalently, dimensionless
$(k^\prime)^\mu$ equals dimensional $(k^\prime)^\mu$ multiplied by
$\lambdabar_C$. 
The choice of the normalization coefficient, 
$$
N_p=\frac{\omega^\prime V}{2\pi\hbar c\lambdabar_C},
$$ 
corresponds to a
single photon in the volume, $V$.

The emission probability,  $dW$,  
is given by an integral  
over $\Delta^4x$:
\begin{equation}\label{eq:probab}
dW=\frac{\alpha L_fL_p}{\hbar c}
\left|\int{\hat{\psi}_f(\gamma\cdot (A^\prime)^*)\psi_idx^0dx^1dx^2dx^3}\right|^2.
\end{equation} 
Here
\begin{equation}\label{eq:firstphase}
L_p=\frac{Vd^3{\bf k}^\prime}{(2\pi\lambdabar_C)^3}=
\frac{\hbar c N_p d^2{\bf k}^\prime_\perp d(k\cdot
k^\prime)}{(2\pi\lambdabar_C)^2(k\cdot k^\prime)}
\end{equation} 
is the number of states for the emitted photon. The transformation of
the phase volume as in Eq.(\ref{eq:firstphase}) is based on the
following Jacobian:
$$
\left(\frac{\partial k^\prime_\|}{\partial(k^\prime\cdot
  k)}\right)_{{\bf k}^\prime_\perp={\rm const}}=\frac{\omega^\prime}{(k^\prime\cdot
  k)},
$$
which is also used below in
many places. A subscript $i,f$ denotes the 
electron in the initial (i) or final (f) state.  
The number of electron states in the presence of the wave field, 
$L_{i,f}$, should be integrated over the volume $V$  
$$
L_{i,f}=\frac{1}{(2\pi\lambdabar_C)^3}\int_V{d^3{\bf p}_{i,f}dV}=
\frac{d(k\cdot p)_{i,f}d^2{\bf p}_{\perp i,f}N_{i,f}}
{2(2\pi)^3\lambdabar_C^3(k\cdot p)_{i,f}}.
$$
\subsection{Conservation laws} 
The integration by $dx^1dx^2dx^3=c\sqrt{2} dtd^2{\bf x}_\perp$  
results in three $\delta-$ functions, expressing the conservation of 
totals of ${\bf p}_\perp$ and $(k\cdot
p)$, for particles in initial and 
final states:
$$
{\bf
  p}_{\perp i}={\bf
  p}_{\perp f}+{\bf
  k}_\perp^\prime,\qquad (k\cdot p_i)=(k\cdot p_f)+
  (k\cdot k^\prime).
$$
Twice integrated with respect to $dx^1$, the probability $dW$ is  
proportional to a long time interval, $\Delta t=\Delta x^1/(c\sqrt{2})$,  
if the boundary condition for the electron wave at $\xi=\xi_-$
is maintained within that long time. On transforming the integral over
$dx^0$ to that over $d\xi$, one can find:
$$
\left|\int{...d^4x}\right|^2 =(2\pi\lambdabar_C)^3 S_\perp c\Delta t\lambdabar\left|\int{...d\xi}\right|^2\times
$$
$$
\times \delta^2({\bf
  p}_{\perp i}-{\bf
  p}_{\perp f}-{\bf
  k}_\perp^\prime)\delta((k\cdot p_i)-(k\cdot p_f)-
  (k\cdot k^\prime)).
$$
To take the large value of $\Delta t$ seems to be the only way to
calculate the integral, however, the emission probability calculated in
this way relates to multiple electrons in the initial state, each of
them locating  between the wave fronts
$\xi=\xi_-$ and $\xi=\xi_+$ during much shorter time, 
\begin{equation}\label{eq:time}
\delta t(\xi_-,\xi_+)=(1/c)\int_{\xi_-}^{\xi^+}{{\cal E}_i(\xi)d\xi_2}/(k\cdot p_i). 
\end{equation}
For a single electron the emission probability becomes: 
$$dW_{fi}(\xi_-,\xi_+)=\delta t dW/\Delta t.$$  
Using $\delta-$ functions
it is easy to integrate Eq.(\ref{eq:probab}) over $d{\bf p}_{\perp f}d(k\cdot
p_f)$:
$$
\frac{dW_{fi}(\xi_-,\xi_+)}{d(k\cdot k^\prime)d^2{\bf k}^\prime_\perp}=
\frac{\alpha \left|
   \int_{\xi_-}^{\xi_+}{T(\xi)\hat{u}(p_f)(\gamma\cdot l^*)u(p_i)d\xi
       }\right|^2}{(4\pi\lambdabar_C)^2(k\cdot k^\prime)(k\cdot
  p_i)(k\cdot p_f)},
$$
where 
$$
T(\xi)=\frac{P_i(\xi)}{P_f(\xi)P_p(\xi)}=
\exp\left[\frac{i\int^\xi{(k^\prime\cdot p_{i}(\xi_2))d\xi_2}}{\lambdabar_C(k\cdot p_{f})}
\right],
$$ 
$P_i(\xi)$ and $P_f(\xi)$ are the electron phase multipliers, $P(\xi)$,
for the electron in initial and final states and 
\begin{equation}\label{eq:cons}
p^\mu_f(\xi)=p^\mu_i(\xi)-(k^\prime)^\mu+\frac{(k^\prime\cdot p_i(\xi))}
{(k\cdot p_i)-(k\cdot k^\prime)}k^\mu . 
\end{equation}

Prior to discussing Eq.(\ref{eq:cons}), return to
Eq.(\ref{eq:QEDpropagator}) and analyze it component-by-component in the
TST. It appears that three of the four components of that equation
describe the conservation of 
$(k\cdot p)$ and ${\bf p}_{\perp 0}={\bf p}_{\perp}+{\bf a}$ 
for electron {\it in the course of its
emission-free motion}. At the same time, yet another component of 
Eq.(\ref{eq:QEDpropagator}), specifically, $p^1$, directed along $k^\mu$, describes the
energy-momentum exchange between the electron and the 1D wave field, maintaining the
identity, $(p\cdot p)=1$. Now turn to Eq.(\ref{eq:cons}). Again, three
of the four components express the conservation of the same variables
{\it in the course of the photon emission}, while
the
 $p^1$ component,
directed along $k^\mu$ describes the absorption of energy and momentum
from the wave field in the course of the photon emission. Note, that in
the case of a strong field, the energy absorbed from field is not an
integer number of quanta, and that for short non-harmonic field it is
not even  
a constant, but a function of the local field.
\subsection{Calculation of the matrix element}
To 
calculate the matrix element, one can re-write it as the 
double integral over $d\xi d\xi_1$ and then reduce the matrices 
$u(p_{i,f}(\xi))\otimes \hat{u}(p_{i,f}(\xi_1))$ in the integrand to the polarization 
matrices of the electron at $\xi$ or at $\xi_1$ using 
Eq.(\ref{eq:QEDpropagator}). These standard manipulations with the Dirac
matrices are omitted here. Although in a strong wave 
electrons may be polarized (see \cite{Omori}),  
in the present work the emission probability is assumed to be averaged
over the electron initial polarizations and summed over its
final polarizations. The ultimate result of these derivations is as follows: 
\begin{equation}\label{eq:dWfi}
\frac{dW_{fi}}{d(k\cdot k^\prime)d^2{\bf k}^\prime_\perp}=
\frac{\alpha \int_{\xi_-}^{\xi_+} {\int_{\xi_-}^{\xi_+}{T(\xi)T(-\xi_1)]D
d\xi d\xi_1}}}{(2\pi\lambdabar_C)^2
(k\cdot k^\prime)(k\cdot p_i)(k\cdot p_f)},
\end{equation}
where
$$
D=
(l^*\cdot p_{i}(\xi_1)) (l\cdot p_i(\xi))-
\frac{\left[(p(\xi)\cdot p(\xi_1))-1\right](1-C_{fi})^2
}{4 C_{fi}}
$$
and
$$
C_{fi}=\frac{(k\cdot p_f)}{(k\cdot p_i)}=1-\frac{\lambdabar_C(k\cdot
  k^\prime)}{(k\cdot p_i)}\le1
$$
is a recoil parameter which characterizes the reduction in the photon
momentum due to emission.

The matrix element may 
also
be 
summed, if desired, for two possible directions of the polarization
vector. The second term in the integrand is simply multiplied by two, while in 
the first one 
the negative of the metric tensor 
 should be substituted for
the product of the polarization vectors 
(see \S8 in \cite{lp}), so that  
$-\left(p_i(\xi)\cdot p_i(\xi_1)\right)$
substitutes for 
$(l^*\cdot p_{i}(\xi_1)) (l\cdot p_i(\xi))$. 
The latter may be transformed using
Eq.(\ref{eq:pp}), thus,  giving:
$$
\sum_{l}{D}=-
\left( 
\frac{\left[(p(\xi)\cdot p(\xi_1))-1\right]\left(1+C_{fi}^2\right)
}{2 C_{fi}}+1\right).
$$
\subsection{Vector amplitude of emission in QED case}
Now 
moving to 
connection between the obtained
result, 
on 
one hand, and the way the high-frequency emission is
treated in the framework of classical theory, on the other. To
facilitate the comparison, both here and in Section II the photon
frequency and wave vector, $\omega^\prime$
and ${\bf k}^\prime$,  are not dimensionless. 
It appears that the QED result obtained above can be reformulated in 
a form similar to Eq.(\ref{eq:Jackson}). Using the
following relationships between the differentials:
$$
dt={\cal E}d\tau=\frac{{\cal E}}{c} \frac{d\xi}{(k\cdot p)},\qquad
dR_{\rm QED}=\hbar\omega^\prime dW_{fi},
$$
$$
\frac{(\omega^\prime)^2d\omega^\prime d{\bf n}}{c^3}=d^3{\bf
  k}^\prime=\frac{\omega^\prime d^2{\bf k}^\prime_\perp d(k\cdot
  k^\prime)}{c(k\cdot k^\prime)},
$$
one can reduce Eq.(\ref{eq:dWfi}) for the {\it polarized} part of the
emission to the same form as that of Eq.(\ref{eq:Jackson}):
$$
\frac{dR_{\rm QED}^{\rm pol}}
{d\omega^\prime d{\bf n}}=\frac{(\omega^\prime)^2}{4\pi^2c}\left|({\bf
  A}_{\rm QED}(\omega^\prime)\cdot {\bf l}^*)\right|^2,
$$
where
$$
{\bf A}_{\rm QED}(\omega^\prime)=\sqrt{\frac{e^2}{C_{fi}}}
\times
$$
$$
\times
\int_{t_-}^{t_+}{\frac{{\bf
    v}(t)}c\exp\left\{\frac{i\omega^\prime}{C_{fi}}[t-\frac{({\bf n}\cdot{\bf
        r}(t))}c]\right\}dt}, 
$$
where $t_-,t_+$ are the time instants when the electron crosses the
hypersurfaces $\xi=\xi_-$ and $\xi=\xi_+$ correspondingly. In the
considered 
strong field case 
the finite integration limits are
admissible as long as the integration span well exceeds the formation time. Therefore, 
$$
{\bf A}_{\rm QED}(\omega^\prime)=\sqrt{
\frac{1}{C_{fi}}} 
{\bf A}_{\rm cl}\left(
\frac{\omega^\prime}{C_{fi}}\right),
$$
\begin{equation}\label{eq:pol}
\frac{dI_{\rm QED}^{\rm pol}(\omega^\prime)}
{d\omega^\prime d{\bf n}}=C_{fi}\frac{dI_{\rm cl}(\omega^\prime/C_{fi})}
{d\omega^\prime d{\bf n}}.
\end{equation}
Thus, the QED effect on the emission from an electron  
reduces to
the classical 
electric-dipole emission from 
a moving charge in 
an electromagnetic field with a very simple rule to transform the polarized
emission intensity and polarized emission amplitude,
which accounts for 
the recoil effect. 

However, the electron in the QED-strong field emits not only as an
electric charge, because it also possesses a magnetic moment
associated with its spin. In Eq.(\ref{eq:dWfi}) 
a depolarized
contribution to the emission is present. This contribution may be related to the 
electron spin, which is 
{\it assumed} 
to be depolarized. The depolarized emission
energy  related to 
the interval of frequency and to the element of solid angle and summed over two
polarization 
directions (i.e., multiplied by two) equals:
$$
\frac{dR^{\rm depol}_{\rm
  QED}}{d\omega^\prime d{\bf
    n}}=
\frac{(\omega^\prime)^2}{4\pi^2c}\frac{\left(1-C_{fi}\right)^2}{2C_{fi}}
\times
$$
$$
\times\left\{\sum_{\bf l}\left|({\bf
  A}_{\rm QED}(\omega^\prime)\cdot {\bf l}^*)\right|^2+\left|\varphi_{\rm QED}\right|^2\right\},
$$ 
where
$$
\varphi_{\rm QED}=\sqrt{\frac{e^2}{C_{fi}}}
\int_{t_-}^{t_+}{
\exp\left\{\frac{i\omega^\prime}{C_{fi}}[t-\frac{({\bf n}\cdot{\bf
        r}(t))}c]\right\}\frac{dt}{{\cal E}(t)}}
$$
is a scalar amplitude of emission, introduced in 
a way similar to
that for the vector amplitude. After 
derivations analogous to those 
of Section II,
the radiation loss rate due to the electron magnetic moment reads:
\begin{equation}\label{eq:depol}
\frac{dI^{\rm depol}_{\rm
  QED}}{d\omega^\prime}=\frac{I_{\rm
    cl}(\tau)}{\omega_c}
\frac{9\sqrt{3}}{8\pi}\left(1-C_{fi}\right)^2
\frac{r_0}{C_{fi}}K_{2/3}\left(\frac{r_0}{C_{fi}}\right).
\end{equation}
Thus, the QED effect in the emission from 
an
electron in a strong
electromagnetic field
reduces to 
a downshift in frequency accompanied by 
an extra
contribution from the magnetic moment of 
electron. Note a general
character of these conclusions: only in the recoil parameter, $C_{fi}$, 
is there a direct dependence on the 1D wave vector. This dependence can
be also excluded, because, 
for the photons emitted along the direction
of the particle motion, 
the following approximation is valid:
$$
C_{fi}\approx
1-\frac{\hbar\omega^\prime}{{\cal E}m_ec^2}=1-\chi r_0,\qquad
\frac{r_0}{C_{fi}}=r_\chi=\frac{r_0}{1-\chi r_0},
$$ 
so that in QED-strong fields the emission spectrum is also universal
and may be parameterized with the sole parameter, $I_{\rm cl}$. 

Combining Eqs.(\ref{eq:thetafforwave},\ref{eq:time},\ref{eq:Deltat}) one
can derive the condition
\begin{equation}\label{eq:bounds}
\xi_+-\xi_-\gg\frac{2}{|d{\bf a}/d\xi|}.
\end{equation}
Under this condition, the time interval within which the emitting
electron locates between the wave fronts $\xi=\xi_-$ and $\xi=\xi_+$
much exceeds the formation time of emission 
validating the above
considerations.

With these results the scheme to account for the 
high-frequency
emission as
outlined in Section II may be easily extended to 3D QED-strong fields. 
However, the radiation back reaction needs to be incorporated for this
scheme to be consistent. 
\section{Radiation and its back-reaction}
\begin{figure}
\includegraphics[scale=0.38]{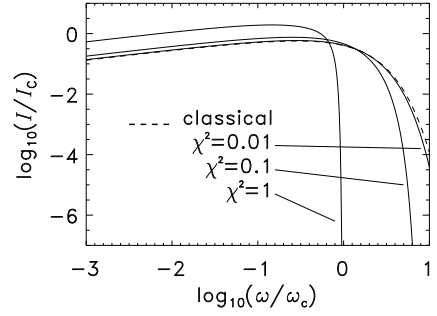}
\caption{Emission spectra for various values of $\chi$.
}
\label{fig_2}
\end{figure}
\begin{figure}
\includegraphics[scale=0.38]{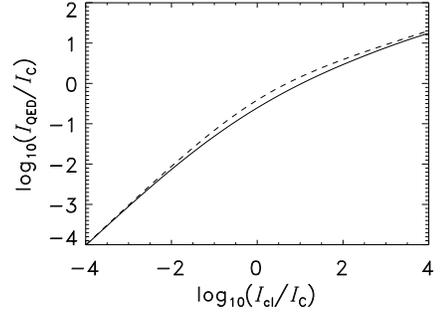}
\caption{Emitted radiation power in the QED approach {\it vs} classical 
(solid); an interpolation formula $I_{\rm QED}=I_{\rm cl}/(1+1.04\sqrt{I_{\rm cl}/I_C})^{4/3}$ (dashed).}
\label{fig_3}
\end{figure}
Unless the field is QED-strong the radiation back-reaction in
a
relativistically strong laser wave may or may not be significant. The
condition for the field to be radiation-dominant (see, e.g.,
Ref.\cite{kogaetal}) is formulated in terms of the ratio between the 
magnitudes of the Lorentz force and 
the radiation force, which
become comparable at intensities $J\sim
10^{23}{\rm W/cm}^2$. For an electron moving toward
the laser wave,
the radiation force 
starts to dominate at a lower wave intensity,
depending on the electron energy \cite{kogaetal}. The radiation
back-reaction 
decelerates 
such 
an
electron, the effect being more 
pronounced for longer laser pulses \cite{our}. As the result, 
at intensities $J\sim 10^{22}{\rm W/cm}^2$ the radiation back-reaction
drastically changes the character of the laser pulse interaction with
dense plasmas and $\gamma$-ray emission becomes a leading mechanism of
the laser energy losses \cite{FI}.

In QED-strong 
fields
the radiation back-reaction is {\it always}
significant,
as long as at
each photon emission the electron looses a noticeable
fraction of its momentum and energy. 
The matter of principle is also a
consistency of the perturbation theory of emission in QED-strong 
fields. Within the framework of classical theory the momentum-energy 
change resulting from the radiation back-reaction should be small in
some sense, to properly approximate the radiation force (see 
\cite{ll},\cite{jack} as well as the considerations relating to the
estimate as in Eq.(\ref{eq:Stepansratio})). In QED-strong 
fields this
change cannot be claimed to be small, but the probability of emission can be! 
Specifically, the difference, $\xi_+-\xi_-$, should be small enough, so
that the probability of emission within the time interval 
of Eq.(\ref{eq:time}) should be much less (or at least less) 
than unity:
\begin{equation}\label{eq:upperbound}
\int{\frac{dW_{fi}}{d{\bf k}^\prime_\perp d(k^\prime\cdot k)}d{\bf k}^\prime_\perp d(k^\prime\cdot k)}\ll1.  
\end{equation}
\subsection{Emission probability and radiation loss rate}
The derivations performed in Section II for the radiation loss rate,
namely, the approximation of the angular distribution with the Dirac
function and approximation of the frequency spectrum with the MacDonald
functions may be directly applied to the emission probability. 

On developing the dot-product, $(k^\prime\cdot p_i)$, in 
$T(\xi)$ in the TST  metric, $G^{\mu\nu}$, one can find: 
$$
T(\xi)T(-\xi_1)=\exp\left[i(T_1+T_2)\right], 
$$
where 
$$T_1=\frac{(k\cdot p_i)}{2\lambdabar_C(k\cdot k^\prime)(k\cdot p_f)}\left(\frac{(k\cdot k^\prime)}{(k\cdot p_i)}\left<{\bf p}_{\perp i}\right>-
{\bf k}^\prime_{\perp}\right)^2(\xi-\xi_1),
$$
$$
T_2=\frac{ (k\cdot k^\prime) \left\{(\xi-\xi_1)+\int_{\xi_1}^\xi{
\left[{\bf a}(\xi_2)-\left<{\bf
    a}\right>\right]^2d\xi_2}\right\}}{2\lambdabar_C(k\cdot p_i)(k\cdot p_f)}, 
$$
$$
\left<{\bf a}\right>=\frac{\int_{\xi_1}^\xi{{\bf
      a}d\xi_2}}{\xi-\xi_1},\qquad \left<{\bf p}_{\perp i}\right>={\bf
  p}_{\perp 0 i}-\left<{\bf a}\right>.$$ 
Integration over $d^2{\bf
  k}^\prime_\perp$ 
then gives: 
$$
\frac{dW_{fi}(\xi_-,\xi_+)}{d(k\cdot k^\prime)}=
\frac{\alpha \int_{\xi_-}^{\xi_+}
  {\int_{\xi_-}^{\xi_+}{\frac{i\exp(iT_2)}{\xi-\xi_1}\sum_{l}{D(\xi,\xi_1)}
d\xi d\xi_1}}}{2\pi\lambdabar_C(k\cdot p_i)^2}.
$$ 
In  
strong fields 
the following estimates may be applied:   
$$
(k\cdot k^\prime)\sim \lambdabar_C(k\cdot p_i)^2\left|\frac{d{\bf a}}{d\xi}\right|,\qquad
dW_{fi}\sim\alpha\left|(\xi_+ - \xi_-)\frac{d{\bf a}}{d\xi}\right|.$$ 
Now the bounds for $\xi_+-\xi_-$ can be {\it consistently} introduced: 
\begin{equation}\label{consistency}
\left|d{\bf a}/d\xi\right|^{-1}\ll\xi_+-\xi_-\ll
\min\left(\alpha^{-1}\left|d{\bf a}/d\xi\right|^{-1},1\right).
\end{equation}
Under these bounds, first, the condition (\ref{eq:bounds}) is
satisfied. 
Therefore, the time interval (\ref{eq:time}) is much greater than the
formation time and
the emission probability is linear in $\xi_+-\xi_-$: 
$$dW_{fi}(\xi_-,\xi_+)=(dW/d\xi)(\xi_+-\xi_-).$$ 
Second, the emission probability satisfies the 
condition (\ref{eq:upperbound}). Therefore,  
perturbation theory is applicable. In addition, 
the emission probability can be expressed in terms of the local 
electric field. Note, that consistency in
(\ref{consistency}) is ensured in relativistically strong
electromagnetic fields as long as $\alpha\ll1$, with no restriction on
the magnitude of the electromagnetic field experienced by an electron.

Under the condition (\ref{consistency}) the probability may be expressed
in terms of McDonald functions:
\begin{equation}\label{eq:probabf}
\frac{dW_{fi}}{dr_0d\xi}=
\frac{\alpha \chi\left(\int_{r_\chi}^\infty{K_{\frac53}(y)dy}+r_0r_\chi\chi^2
K_{\frac23}(r_\chi)\right)}{\sqrt{3}\pi\lambdabar_C(k\cdot p_i)},
\end{equation}
$$ 
r_0=\frac{(k\cdot
  k^\prime)}{\chi(k\cdot p_i)},\qquad
\chi=\frac32(k\cdot p_i)\left|\frac{d{\bf a}}{d\xi}\right|\lambdabar_C=\sqrt{\frac{I_{\rm cl}}{I_C}}.
$$
Probability (similar to that found in \cite{nr}), is expressed in terms of functions of $r_0$, and 
related to interval of
$dr_0$. The way to introduce $r_0$ and $\chi$ looks different from that
adopted in Eqs.(\ref{eq:classicspectrum},\ref{eq:omegaccl}), however, the
difference is negligible as long as ${\omega^\prime}/{{\cal
    E}_i}\approx {(k\cdot k^\prime)}/{(k\cdot p_i)}$   
for 
collinear $k^\prime$ and $p_i$.
The momentum of the emitted radiation, related to the interval of the electron 
{\it proper} time, may be found from Eqs.(\ref{eq:time},\ref{eq:probabf}):
\begin{eqnarray}
\frac{dp_{\rm rad}}{d\tau}=
\int{k^\prime \frac{c(k\cdot p_i)dW_{fi}}{ d(k\cdot k^\prime)d^2{\bf k}_\perp d\xi}d(k\cdot k^\prime)d^2{\bf k}_\perp}=\nonumber\\
=[p+k\ O((k\cdot p_i)^{-1})]\int{c(k^\prime\cdot k) \frac{dW}{dr_0 d\xi}dr_0}.\label{eq:prad}
\end{eqnarray}
As with other 4-momenta, $p_{\rm rad}$ is related to $m_ec$. To prove the 4-vector relationship (\ref{eq:prad}),  
its components in the TST metric 
should be integrated 
over ${\bf k}_\perp$ using the symmetry of $T_1$. The small term,
$O(1/(kp_i))$, arises from the electron rest mass energy  
and from the small 
($\sim 1/|p_\perp|$) but finite width of the photon angular distribution. 
In neglecting this term:
$$\frac{dp_{\rm rad}}{d\tau}=p_i\frac1{m_ec^2}\int{ \frac{dI_{\rm QED}}{dr_0}dr_0},$$
$$I_{\rm QED}=m_ec^2\int{c(k\cdot
  k^\prime)\frac{dW_{fi}}{d\xi dr_0}dr_0}$$ 
being the radiation loss rate.  
The photon energy spectrum, $dI_{\rm QED}/dr_0$,  is described as 
a
function only of
the random {\it scalar}, $r_0$, using only the parameter, $\chi$ (see Fig.\ref{fig_2}). 
The latter may be parameterized in terms of the radiation loss rate, evaluated within 
the framework of classical theory (see Eq.(\ref{eq:chifirst}) and Fig.\ref{fig_3}). The 
expressions for $q(I_{\rm cl})=I_{\rm QED}/I_{\rm cl}$ and for the 
normalized spectrum function, $Q(r_0,\chi)$,  coincide with formulae known from the 
gyrosynchrotron emission theory (see \S90 in \cite{lp}): 
$$
q=\frac{9
\sqrt{3}}{8\pi}\int_0^\infty{dr_0r_0\left(\int_{r_\chi}^\infty{K_{\frac53}(y)dy}+r_0r_\chi\chi^2
K_{\frac23}(r_\chi)\right)},
$$
$$
Q(r_0,\chi)=\frac{9\sqrt{3} r_0}{8\pi q}\left(\int_{r_\chi}^\infty{K_{\frac53}(y)dy}+
r_0r_\chi\chi^2
K_{\frac23}(r_\chi)\right).
$$
As mentioned above,
$$
\frac{dI_{\rm QED}}{dr_0}=\frac{dI^{pol}_{\rm QED}}{dr_0}+\frac{dI^{depol}_{\rm QED}}{dr_0},
$$
where polarized and non-polarized contributions are given by
Eqs.(\ref{eq:pol}-\ref{eq:depol})
\subsection{Radiation back-reaction: radiation force approximation} 
\begin{figure}
\includegraphics[scale=0.4]{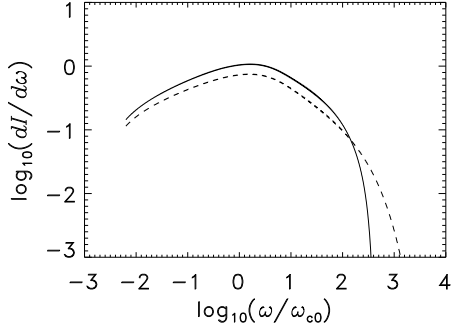}
\caption{The emission spectrum for 600 MeV electrons interacting with 
30-fs laser pulses of intensity $2\cdot 10^{22} W/cm^2 $: with (solid) or 
without (dashed) accounting for the QED effects. Here $\hbar\omega_{c0}\approx 1.1$ MeV for $\lambda=0.8\mu$m.
}
\label{fig_4}
\end{figure}
While emitting a photon, an electron also acquires 
4-momentum from the external field (see Eq.(\ref{eq:cons})): 
$$
dp^\mu_F=\frac{(k^\prime\cdot p_i(\xi))}
{(k\cdot p_i)-(k\cdot k^\prime)}k^\mu.
$$
The interaction with the field ensures that the {\it total} effect of
emission on the electron not to break the entity $(p_f\cdot p_f)=1$. 
As long as the angular distribution of emission is approximated with the Dirac
function,  
the expression for  $dp^\mu_F$ needs to be corrected to ensure 
exact momentum-energy conservation with approximated momentum of
radiation. The 
choices of near-unity correction coefficients in $dp_F$ are somewhat different 
in the cases $\chi\le 1$ and $\chi\gg 1$. 

For moderate values of $\chi\le 1$ the 
{\it radiation force}, $(dp_F-dp_{\rm rad})/d\tau$, may be
introduced. In this approximation it is admitted that the change in the 
electron momentum within the 
infinitesimal time interval is 
also infinitesimal.
This 
'Newton's law' approximation is pertinent to classical physics and
it both ignores the point that the change in the electron momentum at 
$\chi\sim1$ is essentially finite because of the finite momentum of
emitted photon and breaks the low bound on the time 
interval presented in (\ref{eq:bounds}). The approximation, however, is highly efficient and allows 
one
to avoid time-consuming statistical simulations. The approximation error
tends to zero as $\chi\rightarrow0$, however, it is
not huge at $\chi\sim1$ and even at $\chi=10$. The latter can be seen
from  
Fig.\ref{fig_5} given below 
in which the average  
relative change in the
electron energy in the course of
single-photon emission is presented (assumed to be negligible
within the radiation force approximation).  

Within the radiation force approximation the best correction 
is 
$dp^\mu_F
\approx k^\mu 
(k^\prime\cdot p_i)/(k\cdot p_i)$.
The total radiation force 
may now be found
by integrating $dp_F$
over  
$d(k^\prime\cdot k)$: 
\begin{equation}\label{eq:radf}
\frac{d(p^\mu_f-p^\mu_i)}{d\tau}=\left(k^\mu\frac{(p_i\cdot p_i)}{(k\cdot p_i)}-p^\mu_i\right)\frac{I_{\rm QED}}{m_ec^2}.
\end{equation} 
The radiation force maintains the abovementioned entity as long as
$(p_i\cdot d(p_f-p_i)/d\tau)=0$. 
Eq.(\ref{eq:radf}) is presented in 
a form which is applicable 
both with dimensionless or with dimensional momenta. 

In \cite{mine},\cite{our} it was mentioned,  that QED is not compatible 
with the traditional approach to the radiation force in classical electrodynamics and 
an alternative equation of motion for a radiating 
electron was suggested:
\begin{equation}\label{eq:our}
\frac{dp^\mu}{d\tau}=
\Omega^{\mu\nu}p_\nu-
\frac{I_{\rm QED}}{m_ec^2}p^\mu+\tau_0\frac{I_{\rm QED}}{I_{\rm cl}}
\Omega^{\mu\nu}\Omega_{\nu\beta}p^\beta,
\end{equation} 
where $\Omega^{\mu\nu}=eF^{\mu\nu}/(m_ec)$, $F^{\mu\nu}$ is the field
tensor and 
$$\tau_0=2e^2/(3m_ec^3)=(2/3)\alpha\lambdabar_C/c.$$ 
In the 1D plane wave the particular expression for the radiation force
can be found using the following equation:
$$\tau_0\Omega^{\mu\nu}\Omega_{\nu\beta}p^\beta=k^\mu\frac{(p\cdot p)I_{\rm
  cl}}{m_ec^2(k\cdot p)}.$$ 
With this account, the radiation force in Eq.(\ref{eq:our})
is the same as its QED formulation in Eq.(\ref{eq:radf}). This proves that the 
earlier derived Eq.(\ref{eq:our}) has a wide range of applicability including 
electron quasi-classical 
motion in QED-strong fields and in the particular case of 
1D wave
fields it can be directly derived from the QED principles. Note, that
the efforts to derive the radiation force from quantum mechanics were
applied many times (see \cite{moniz}, the most
convincing approach which gives the equation 
quite similar to
(\ref{eq:our}) may be found in \cite{neil}). However, for the first
time the derivation from QED side is provided, with the resulting
equation being different from those given in textbooks \cite{ll},\cite{jack}.

The way to solve Eq.(\ref{eq:our}) within the PIC scheme and integrate the 
emission is described in \cite{our}. In Fig.\ref{fig_4} we show the numerical result for an electron interacting with a
laser pulse. We see that the QED effects essentially modify the radiation spectrum even 
with laser intensities which are already achieved.  

\subsection{Radiation back-reaction: Monte-Carlo approach}
The radiation force approximation does not fully account for the
statistical character of the emission process at $\chi\ge
1$. Specifically, we mentioned above that in the 'Newton's law'
approximation, the force, ${\bf
  f}$, provides only the {\it infinitesimal} change in the electron momentum, 
$\Delta{\bf p}={\bf f}\Delta t\rightarrow 0$ over an infinitely short time interval,
as $\Delta t\rightarrow 0$. For radiation
processes in QED-strong fields this point is in contradiction with
a small {\it probability}, $\Delta t\cdot dW_{fi}/dt\rightarrow0$, for
an electron to acquire a {\it finite} change in momentum,
$|\delta{\bf p}|\sim|{\bf p}|$, in the course of emission.
\begin{figure}
\includegraphics[scale=0.4]{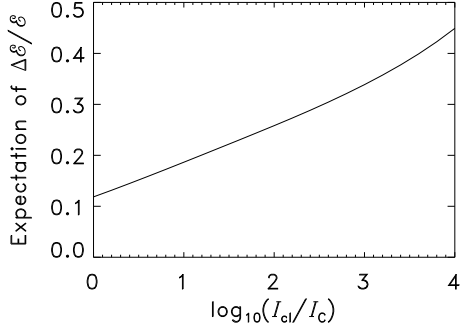}
\caption{Expectation of the emitted photon energy. In dimensional units 
$\Delta{\cal E}=\hbar\omega^\prime$ and ${\cal E}$ is the dimensional
  energy of an electron prior to emission.}
\label{fig_5}
\end{figure}
   
A more quantitative, though more cumbersome, description may be
achieved within the QED Monte-Carlo approach. It is convenient to
relate the emission probability to an interval of proper time,
$\Delta\tau=\Delta t/{\cal E}_i$. From (\ref{eq:probabf}) it follows that:
\begin{equation}\label{eq:probabtau}
\frac{dW_{fi}}{dr_0d\tau}=\frac{I_{\rm QED}}{m_ec^2}\frac{Q(r_0,\chi)}{\chi r_0}.
\end{equation}
(Note, that on multiplying Eq.(\ref{eq:probabtau}) by
$m_ec^2(\omega^\prime/{\cal E}_i)\approx m_ec^2\chi r_0$, one obtains again
the formula for spectral distribution of energy emitted per interval of time.)
As long as Eq.(\ref{eq:probabtau}) for the differential probability is available,
one can find the expected energy of the emitted photon:
$$\frac{1}{{\cal
E}_i}\left<\omega^\prime\right>=
\frac\chi{\int_0^{1/\chi}{Q(r_0,\chi)(dr_0/r_0)}}. 
$$
A plot of the expectation for the ratio, $<\omega^\prime/{\cal E}_i>$, vs $\chi$ is
given in Fig.\ref{fig_5}, with energy of emitted photons being denoted
as $\Delta{\cal E}$. 
The total probability of emission within the interval of proper time is
given by a complete integral of probability:
$$
W_{fi}=
\Delta\tau
\int{\frac{dW_{fi}}{dr_0}d\tau dr_0}=\Delta\tau\frac{I_{\rm QED}}{m_ec^2}\left<\frac{\omega^\prime}{{\cal
E}_i}\right>^{-1}.
$$
Both within the QED perturbation theory and within the Monte-Carlo
scheme $W$ is assumed to be small $W<1$. The probability of no emission 
equals $1-W\ge0$. The partial probability, $W_{fi}
(\omega^\prime<\omega^\prime_0)$, for the emission with the photon
energy 
not exceeding the given value,
$\omega^\prime_0$, is given by the incomplete probability integral:
$$
W_{fi}
(\omega^\prime<\omega^\prime_0)=
W_{fi}\left<\frac{\omega^\prime}{\cal E}\right>\int^{
\omega^\prime/({\cal E}_i\chi)}{
Q(r_0,\chi)\frac{dr_0}
{\chi r_0}
}.
$$
Therefore, for a given interval of proper time and calculated
$\chi$, $\left<\omega^\prime\right>/{\cal E}$ and $W_{fi}<1$,
the expression of the only scalar to gamble,
$\omega^\prime/{\cal E}_i$, in terms of a random number, 
$0\le {\rm rnd}<1$, is implicitly 
given by an integral equation as follows: 
$$
\int^{
\omega^\prime_0/({\cal E}_i\chi)}_0{
Q(r_0,\chi)\frac{dr_0}
{\chi r_0}
}=\frac{\rm rnd}{W_{fi}}\left<\frac{\omega^\prime}{ {\cal E}_i }\right>^{-1},
$$
if the gambled value of ${\rm rnd}$ does not exceed
$W_{fi}$: $0\le{\rm rnd}\le W_{fi}$. Otherwise, i.e. if $W_{fi}<{\rm rnd}\le1$, the emission
within this interval does not occur.

Once the value of $\omega^\prime/{\cal E}$ is found, the change in the
electron 4-momentum due to 
single photon emission during the time
interval, $\Delta\tau$, may be determined as follows:
$$
p^\mu_f-p^\mu_i=\left\{k^\mu\frac{(p_i\cdot p_i)[1-\omega^\prime/(2{\cal
  E})]}{(k\cdot
  p_i)[1-\omega^\prime/{\cal E}]}-p^\mu_i\right\}\frac{\omega^\prime}{\cal E}.
$$
It is easy to see that the identity $(p_f\cdot p_f)=1$ is maintained. To
achieve this, a correction factor, $1-\omega^\prime/(2{\cal E})$ is
applied to the momentum exchange with the wave field as present in Eq.(\ref{eq:cons}).

The implementation of this method for 3D realistic laser fields together
with simulation results and an account for pair production will be
described in detail in a forthcoming publication.
\section{Conclusion}
QED-strong fields in the focus of an ultra-bright laser may be realized,
if desired, using the technologies which already exist. In any case,
these effects will come into power when 
laser-plasma interactions 
are explored with the next generation of lasers.

It is 
demonstrated that 
electron motion in very strong laser
fields with pronounced QED effects may be 
successfully described within the radiation force approximation. The necessary 
corrections in the radiation force and the emission spectra to account for the QED 
effects are parameterized by the {\it sole} parameter, $I_{\rm cl}$.

{\bf We acknowledge} an invaluable help and fruitful advices we
received from S.S. Bulanov. V.T. Tikhonchuk kindly pointed out some
effects we missed. We are grateful to J. Rafelski, R.F. O'Connel and
V. Hnizdo for critical comments and to V. Tenishev for discussing the
Monte-Carlo method. The work of one of the authors (I.S.) on high energy
density physics is supported by the DOE NNSA under the Predictive
Science Academic Alliances Program by grant DE-FC52-08NA28616.  

\end{document}